\documentclass{article}
\usepackage{alltt}
\usepackage[all]{xy}
\usepackage{latexsym,amssymb,amsmath,amsfonts, amsthm, xcolor}

\def\vbar{\mathchoice{\vrule height6.3ptdepth-.5ptwidth.8pt\kern-.8pt}
  {\vrule height6.3ptdepth-.5ptwidth.8pt\kern-.8pt}
  {\vrule height4.1ptdepth-.35ptwidth.6pt\kern-.6pt}
  {\vrule height3.1ptdepth-.25ptwidth.5pt\kern-.5pt}}
\def\fudge{\mathchoice{}{}{\mkern.5mu}{\mkern.8mu}}
\def\bbc#1#2{{\rm \mkern#2mu\vbar\mkern-#2mu#1}}
\def\bbb#1{{\rm I\mkern-3.5mu #1}}
\def\bba#1#2{{\rm #1\mkern-#2mu\fudge #1}}
\def\bb#1{{\count4=`#1 \advance\count4by-64 \ifcase\count4\or\bba A{11.5}\or
  \bbb B\or\bbc C{5}\or\bbb D\or\bbb E\or\bbb F \or\bbc G{5}\or\bbb H\or
  \bbb I\or\bbc J{3}\or\bbb K\or\bbb L \or\bbb M\or\bbb N\or\bbc O{5} \or
  \bbb P\or\bbc Q{5}\or\bbb R\or\bbc S{4.2}\or\bba T{10.5}\or\bbc U{5}\or
  \bba V{12}\or\bba W{16.5}\or\bba X{11}\or\bba Y{11.7}\or\bba Z{7.5}\fi}}

\newcommand{\RR}{\mbox{${\rm \:  R\!\!\!\! I
\;\;}$}}

\newcommand{\vs}{\vspace{0.25cm}}

\newtheorem{theorem}{Theorem}
\newtheorem{itlemma}{Lemma}[section]
\newtheorem{itproposition}[itlemma]{Proposition}
\newtheorem{itcorollary}[itlemma]{Corollary}
\newtheorem{itremark}[itlemma]{Remark}
\newtheorem{itremarks}[itlemma]{Remarks}
\newtheorem{itdefinition}[itlemma]{Definition}
\newtheorem{itexample}[itlemma]{Example}

\newenvironment{lemma}{\begin{itlemma}\rm}{\end{itlemma}} 
\newenvironment{remark}{\begin{itremark}\rm}{\end{itremark}} 
\newenvironment{remarks}{\begin{itremarks} \rm}{\end{itremarks}}
\newenvironment{corollary}{\begin{itcorollary}\rm}{\end{itcorollary}}
\newenvironment{proposition}{\begin{itproposition}\rm}{\end{itproposition}}
\newenvironment{definition}{\begin{itdefinition}\rm}{\end{itdefinition}}
\newenvironment{example}{\begin{itexample}\rm}{\end{itexample}}
\newenvironment{fact}{\noindent {{\bf Fact}}:\ \ }{\hfill \medskip}
\newenvironment{claim}{\noindent {\em Claim}. \ \ }{\hfill \medskip}
\newcommand{\be}[1]{\begin{equation}\label{#1}}
\newcommand{\ee}{\end{equation}}
\newcommand{\bl}[1]{\begin{lemma}\label{#1}}
\newcommand{\br}[1]{\begin{remark}\label{#1}}
\newcommand{\brs}[1]{\begin{remarks}\label{#1}}
\newcommand{\bt}[1]{\begin{theorem}\label{#1}}
\newcommand{\bd}[1]{\begin{definition}\label{#1}}
\newcommand{\bp}[1]{\begin{proposition}\label{#1}}
\newcommand{\bc}[1]{\begin{corollary}\label{#1}}
\newcommand{\bfact}[1]{\begin{fact}\label{#1}}
\newcommand{\bex}[1]{\begin{example}\label{#1}}
\newcommand{\ec}{\end{corollary}}
\newcommand{\efact}{\end{fact}}
\newcommand{\eex}{\end{example}}
\newcommand{\el}{\end{lemma}}
\newcommand{\er}{\end{remark}}
\newcommand{\ers}{\end{remarks}}
\newcommand{\et}{\end{theorem}}
\newcommand{\ed}{\end{definition}}
\newcommand{\ep}{\end{proposition}}
\newcommand{\epr}{\end{proof}}
\newcommand{\bpr}{\begin{proof}}
\newcommand{\bcl}{\begin{claim}}
\newcommand{\ecl}{\end{claim}}

\newcommand{\bi}{\begin{itemize}}
\newcommand{\ei}{\end{itemize}}
\newcommand{\ben}{\begin{enumerate}}
\newcommand{\een}{\end{enumerate}}

\usepackage{graphicx}

\title{\bf \Large{TIME OPTIMAL SIMULTANEOUS CONTROL OF TWO LEVEL QUANTUM SYSTEMS}}

\vs

\vs

\author{Francesca Albertini\thanks{Dipartimento di Matematica,  Universit\`a di Padova, albertin@math.unipd.it}  \, \, \,  and \, Domenico D'Alessandro\thanks{Department of Mathematics, Iowa State University, Ames, Iowa, U.S.A., e-mail:daless@iastate.edu}}

\begin{document}

\maketitle

\begin{abstract}
In this paper, we solve the problem of simultaneously driving  in minimum time  to arbitrary final conditions, $N$ two level quantum systems subject to independent controls. The solution of this problem is obtained via an explicit description of the reachable set of the associated control system on $SU(2)$. The treatment generalizes previous results on the time optimal control of two level quantum systems and suggests that similar techniques could be used to solve the minimum time control problem for a larger class of right invariant systems on Lie groups.

\end{abstract}

\section{Introduction}

In the recent work,  \cite{Noi} \cite{Raf}, the problem of the minimum  time optimal control of a two level quantum system ({\it qubit}) was solved. In the model considered, the system was  subject to a drift and a control field orthogonal to the drift and with bounded norm (cf. section \ref{Rev} for a precise statement of the optimal control problem). Two level quantum systems are of paramount importance in quantum mechanics and, in particular, in  quantum computation (see, e.g., \cite{NC}). In this context, evolution in minimum time  is a natural requirement when it is desired to minimize the effect of the environment or to increase the speed of implementation of a given quantum computation. In fact, the circuit model of quantum computation (cf. \cite{NC}) requires a cascade of simple evolutions on elementary quantum systems which therefore have to occur in very short time in order to maximize the speed of the overall computation. The solution of the minimum time problem given in \cite{Noi} \cite{Raf} is quite simple and explicit, requiring only very elementary numerical work, something very rare for optimal control problems.

When trying to control $N \geq 2$  qubits {\it simultaneously} with independent controls in minimum time, one could argue that the above results could simply be applied $N$ times to obtain  the optimal control. Let us assume, for simplicity of exposition, $N=2$, and let $X_{f1}$ and $X_{f2}$ the desired final condition for system $S_1$ and system $S_2$,  respectively. Let us denote by $T_1$ and $T_2$ the optimal times to reach $X_{f1}$ and $X_{f2}$ for system $S_1$ and $S_2$, respectively. If $T_1=T_2$, then the two optimal controls designed with the techniques of \cite{Noi} \cite{Raf},  will drive the two systems to the desired final conditions, in minimum time. However if $T_1 \not= T_2$, then there is a {\it slow} system and a {\it fast} system and the solution is not simply to slow down the fast system to synchronize it with the slow system. The problem is that for a (bilinear) system with drift, such as the ones considered here, the fact that a certain evolution can be performed  at (minimum) time  $T_1$ does not ensure that the same evolution can be achieved at a later time $T_2 > T_1$. Therefore, this problem requires a careful analysis of the {\it reachable sets} ${\cal R}(T)$,  for the systems under consideration, that is, the set of evolutions or states which can be reached at exactly time $T$.

The goal of this paper is twofold. On one hand we want to solve the above mentioned time optimal and synchronization control problem for $N$ qubits to any desired final condition. On the other hand we want to describe the structure of the reachable sets for the dynamics of two level quantum systems. In doing so we will determine the features of the dynamics of quantum bits which make the time optimal control problem amenable of such a simple solution as described in \cite{Noi}, \cite{Raf}. In fact the underlying geometric structure of the problem, which facilitates its solution, can be found in other problems as well.

The paper is organized as follows. In section \ref{Rev} we describe the problem of minimum time control for a two level quantum system which mathematically amounts to a time optimal control problem for a right invariant bilinear system on the Lie group $SU(2)$. We shall consider and review the main results and the approach of \cite{Noi}, \cite{Raf}. We will see that the time optimal synthesis can be visualized in the unit disk where all the geometric analysis can be performed. A special trajectory called the {\it critical trajectory} plays a special role in that time optimal trajectories loose optimality intersecting this curve. In section \ref{Contin}, we further analyze the optimal synthesis for two level quantum systems by studying the continuity properties of the minimum time function as a function of the bound on the controls or of  the final point. This analysis sheds further light on the time optimal synthesis for this model. It shows that the minimum time function is continuous except for points on the critical trajectory where it presents a right discontinuity. This discontinuity is a manifestation of the fact that the reachable sets for this model do not monotonically grow with time.  The analysis of reachable sets is performed in section \ref{Reachsets}. Here using a change of coordinates we reduce the problem to the  study of the reachable sets for a driftless system. We use monotonicity of the reachable sets in this case and the relation between the geometry of reachable sets and time optimal control trajectories. In this section, we try to present the results in a way that can be applied to more general systems on Lie groups highlighting the features of the system which allow the treatment to go through.  Finally, the description of the reachable sets is used in section \ref{synchro} to give  an algorithm for the design  of the  synchronous time optimal control for $N$ qubits.

\section{The time optimal control problem for a two level quantum system}\label{Rev}

Let us consider the {\it Schr\"odinger operator equation} (see, e.g., \cite{Sakurai}) for a spin $\frac{1}{2}$ particle in a magnetic field with time varying components  (controls) in the $x$ and $y$ direction, $u_x$ and $u_y$. The equation is written as
\be{basicmodel}
\dot X=\tilde \sigma_z X + u_x \tilde \sigma_x X+u_y \tilde \sigma_y X, \qquad X(0)={\bf 1},
\ee
where $X \in SU(2)$, $X(0)={\bf 1}$ the identity,  and $\tilde \sigma_{x,y,z}$ are proportional to the {\it Pauli matrices},  $\sigma_{x,y,z}$, and  form a basis of the Lie algebra $su(2)$. They are defined as
\be{Paulimat}
\tilde \sigma_x:=\frac{i}{2}\sigma_x= \frac{1}{2}\begin{pmatrix} 0 & i \cr i & 0\end{pmatrix}, \qquad
\tilde \sigma_y:=\frac{-i}{2}\sigma_y= \frac{1}{2} \begin{pmatrix} 0 & -1 \cr 1 & 0 \end{pmatrix},
\qquad \tilde \sigma_z:=\frac{i}{2} \sigma_z= \frac{1}{2}\begin{pmatrix} i & 0 \cr 0 & -i\end{pmatrix}.
\ee
The Lie algebra $su(2)$ is equipped with an inner product between  matrices, $\langle \cdot, \cdot \rangle$, defined as $\langle A, B \rangle:=Tr(A B^\dagger)$, so that the associated norm is $\|A\|:=\sqrt{\langle A, A \rangle}$. The coefficient of $\tilde \sigma_z$ in (\ref{basicmodel}) (which is  called {\it Larmor frequency} in NMR applications see, e.g., \cite{Abragam}) is taken equal to $1$, in the appropriate units, without loss of generality, since to this situation we can always reduce ourselves with an appropriate re-scaling and-or reversing  of the time variable and-or a redefinition of the bound on the control\footnote{This is true unless the Larmor frequency is zero in which case we have a driftless system which will be considered in detail in the following (see section \ref{Reachsets}).} (cf. Remark 1.1 in \cite{Noi})

The problem considered in \cite{Noi} and \cite{Raf} is, given a desired final condition $X_f \in SU(2)$, to find the control functions $u_x,u_y$, that steer the state $X$ of system (\ref{basicmodel}) from the identity, ${\bf 1}$, to $X_f$ in minimum time, under  the constraint  that $u_x^2(t)+u_y^2(t) \leq \gamma^2$, for every $t$. We shall later generalize this problem (cf. section \ref{synchro}) to the minimum time {\it simultaneous} control of $N$ two level quantum systems.

The classical approach to the solution of this type of problems is to apply the {\it Pontryagin Maximum Principle PMP} (see, e.g., \cite{Agrachev}, \cite{Flerish}, and,  in particular  \cite{Mikobook}, \cite{Tesi}  for applications to quantum systems.), which gives the necessary conditions for optimality. This results in a set of candidate optimal control functions (more or less explicit) which are typically parametrized by some real values. Then these controls are placed back in the dynamics which is (numerically) integrated. The parameters are then chosen so that the final condition is met and the time of transfer is the minimum one. In essence, the PMP allows one to reduce a search over a space of functions (the controls) to a search over a finite dimensional space (the space of the parameters). In the case of system (\ref{basicmodel}) application of the PMP\footnote{Along with a result showing the non-optimality of singular extremals (cf. \cite{Noi}).} shows that the optimal candidate controls (extremals)  are of the type  \cite{Noi}:
\be{NSE}
u_x=\gamma \sin(\omega \tau + \tilde \phi), \qquad u_y=-\gamma \cos(\omega \tau +\tilde \phi),
\ee
where $\tau$ denotes the time variable and $\omega$ and $\tilde \phi$ are two parameters  (frequency and phase) to be tuned in order to reach the desired final condition while minimizing the time.

 By plugging $u_x$ and $u_y$ in (\ref{NSE}) into (\ref{basicmodel}), the corresponding differential equation {\it can be explicitly integrated}.
The solution is given by
\be{soluzexpli}
X(\tau,\omega,\tilde \phi):=\begin{pmatrix} e^{i \omega t}(\cos(a t)+ i \frac{b}{a} \sin(a t)) & e^{i (\omega t + \tilde \phi)} \frac{\gamma}{a} \sin(a t) \cr  - e^{-i(\omega t + \tilde \phi)} \frac{\gamma}{a} \sin(a t)  & e^{-i \omega t}(\cos(a t)-i \frac{b}{a} \sin(a t)) \end{pmatrix},
\ee
for $t:=\frac{\tau}{2}$, $b:=1-\omega$, $a:=\sqrt{\gamma^2 + b^2}$. Direct inspection of formula (\ref{soluzexpli}) shows that the $\tilde \phi$ only affects the phase of the off-diagonal element. This means that the minimum time only depends on the $(1,1)$ element of the matrix giving the desired final condition.\footnote{This can also be proved without using the explicit form of the optimal controls (cf. Proposition 2.1 in \cite{Noi}).} We can use an arbitrary phase (for example $\tilde \phi=0$) and study the trajectory of the $(1,1)$ element which belongs to the unit disk. Once the frequency $\omega$ corresponding to the time optimal control steering to the desired point of the unit disk has been found, then the phase $\tilde \phi$ is chosen to match the phase of the off diagonal element of the desired final condition.

\begin{figure}[htb]
\centering
\includegraphics[width=0.7\textwidth]{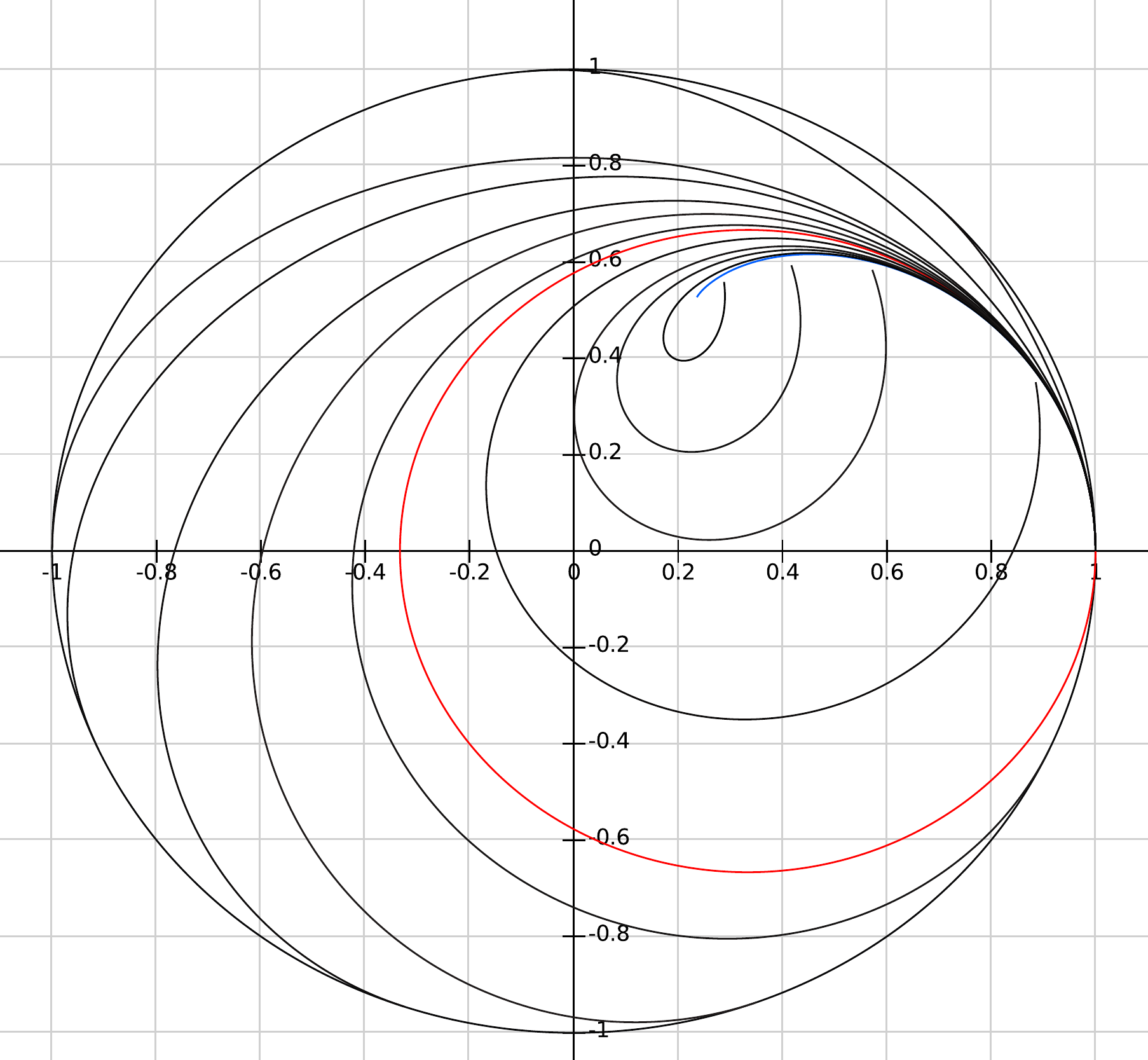}
\caption{Optimal synthesis for $\gamma=\frac{1}{\sqrt{2}}$. Reported here are the separatrix in red corresponding to $\omega=\omega^*=\frac{3}{4}$ and the critical curve in blue corresponding to $\omega=2\omega^*=\omega_c=\frac{3}{2}$. Moreover in black are the optimal trajectories corresponding to $\omega=-3$,  $\omega=-1$, $\omega=0$, $\omega=0.2\omega^*$, $\omega=0.5\omega^*$, $\omega=0.7\omega^*$, $\omega=0.9\omega^*$, outside  the separatrix, and $\omega=1.3\omega^*$, $\omega=1.45\omega^*$, $\omega=1.6\omega^*$, $\omega=1.8\omega^*$.}
\label{Fig1}
\end{figure}

Let $x=x(t)$ and $y=y(t)$ be the real and the imaginary part of  the (1,1) element in  (\ref{soluzexpli}), which, parametrized by $\omega$, is given by:
\be{curvex}
x(t):=x_{\omega}(t)=\cos(\omega t)\cos(at)-\frac{b}{a}\sin(\omega t)\sin(at),
\ee
\be{curvey}
y(t):=y_\omega(t)= \sin(\omega t)\cos(at)+\frac{b}{a}\cos(\omega t)\sin(at).
\ee
In \cite{Noi} the optimal synthesis was described in the unit disk.
This was done for values of $\gamma$, $\frac{1}{\sqrt{3}} \leq \gamma \leq 1$. A typical picture for the optimal trajectories is in Figure \ref{Fig1} to which we shall refer in the following discussion. In general, the values of $\omega$ for which the controls in (\ref{NSE}) are optimal are $-\infty < \omega \leq 1+\gamma^2:=\omega_c$. The limit value $\omega_c:=1+\gamma^2$ called the {\it critical frequency} corresponds to a trajectory, also called the {\bf critical trajectory} which has a cuspid at the time $T_c=\frac{\pi}{2 \gamma \sqrt{1+\gamma^2}}$. The critical trajectory is optimal until time $T_c$ and then looses optimality. It is the curve in blue in Figure \ref{Fig1}. Among the other trajectories, another important one is the one corresponding to $\omega=\omega^*:=\frac{1+\gamma^2}{2}$ which corresponds to a circle (in red in Figure \ref{Fig1}) centered at   $\left( \frac{\gamma^2}{1+\gamma^2}, 0\right)$ and with radius $\frac{1}{1+\gamma^2}$. This trajectory was called the {\it separatrix} since optimal trajectories corresponding to $-\infty < \omega < \omega^*$ entirely lay outside of the region bounded by it until they loose optimality upon reaching the boundary of the unit disk. Also, trajectories with $\omega^* < \omega \leq \omega_c$ remain {\it inside} the separatrix until reaching the critical trajectory and loosing optimality there\footnote{The loss of optimality is of different type at the boundary of the unit disk and on the critical trajectory. At the boundary of the unit disk the trajectory (with corresponding $\omega$) is optimal until an {\it including} the point on the unit disk. However,  for a curve intersecting the critical trajectory, this curve is optimal until {\it but not including}, the point on the critical trajectory.}(sample trajectories are drawn in black in Figure \ref{Fig1}).

Given, the qualitative picture of the optimal synthesis as in Figure \ref{Fig1} it is  straightforward to find the time optimal control to steer to a desired final condition. Let $X_f$ be the desired final condition and $P_f$ the point in the unit disk representing the $(1,1)$ entry of this matrix. Then one finds $\omega$ such that the corresponding trajectory contains $P_f$ (this can be done for example using a bisection algorithm and the graphics of the trajectories). Once $\omega$ has been found one determines the minimum time. This can be done for example solving  a static optimization problem, minimizing (in $t$) the distance of the trajectory with fixed $\omega$ from the point $P_f$. Finally, one determines the phase $\tilde \phi$ in (\ref{NSE})  to match (in (\ref{soluzexpli})) the phase of the off diagonal entry of the desired final condition.

\vs

The assumption  of $\gamma \leq 1$ was used  in \cite{Noi} to guarantee that the minimum time to reach points on the boundary of the unit disk set is an increasing function of the phase of the point. Without this  feature the optimal synthesis becomes more complicated. The assumption of $\frac{1}{\sqrt{3}} \leq \gamma$ was used to render  the synthesis inside the separatrix analytically tractable. As $\gamma \rightarrow 0$ the critical trajectory become longer and longer and resembles a long spiral filling the whole disk bounded by the separatrix. Simulations show that the optimal trajectories follow this trajectory before going around its endpoint (the one corresponding to $T_c=\frac{\pi}{2 \gamma \sqrt{1+\gamma^2}}$) and intersecting the critical trajectory. This behavior observed with numerical simulations is however difficult to describe analytically.

These difficulties were overcome in \cite{Raf} where the author reconsidered the curves (\ref{curvex}), (\ref{curvey}) but this time keeping $t$ fixed and considering $\omega$ as a variable. If this is done, the curves (\ref{curvex}) and (\ref{curvey}) represent, for $\omega$ in a certain interval, part of the boundary of the reachable set at that time $t$. For every $\omega$ in that interval the corresponding point is the endpoint of an optimal trajectory. The points where the curve in (\ref{curvex}) (\ref{curvey}) are not endpoints of optimal trajectories anymore are the ones where the curve corresponding to time $t$, say ${\cal F}_t$\footnote{This curve is called the {\it optimal frontline} in \cite{Raf}.} intersects the one  at time $t+dt$, ${\cal F}_{t+dt}$. The curve of all these intersections with varying $t$ {\it coincides with the critical trajectory above discussed} which is therefore the {\it envelope} of these curves. As $\gamma \rightarrow \infty$ the critical trajectory becomes shorter and shorter and disappears at the limit which corresponds to a driftless system\footnote{Recall that in (\ref{basicmodel}) we have normalized the Larmor frequency to $1$ and the value of $\gamma$ is in fact $\gamma:=\frac{\gamma^{'}}{\omega_0}$ where $\gamma^{'}$ is the `physical' bound on the norm of the control and $\omega_0$ is the value of the Larmor frequency which tends to  zero for a driftless system.} At the other end, when $\gamma \rightarrow 0$ the critical trajectory becomes very long. Besides giving an  interpretation of the critical trajectory, the analysis of \cite{Raf} provides an alternative and general method to find the minimum time control. One consider the frontlines ${\cal F}_t$ with evolving $t$ and bounded on one end by the boundary of the unit disk and on the other end by the critical trajectory (i.e., (\ref{curvex}), (\ref{curvey}) with $\omega=\omega_c$) and look for the smallest $t$ such that the desired final point $P_f$ is in ${\cal F}_t$. This idea will be used later in this paper. We refer to \cite{Raf} for details.

In the next two sections we shall further elaborate on these results and investigate  the geometric nature of the minimum time problem on $SU(2)$ and its relation with the geometry of the reachable sets.

\section{Properties of the minimum time function}\label{Contin}

We now study the continuity  and monotonicity properties of the minimum time function for the above problem for a two level quantum system. This is a function of the final state $X_f \in SU(2)$  fixed and of the bound on the control  $\gamma$. As discussed above  $X_f$ is represented by a point $P_f$ in the unit disc. For any $\gamma>0$, and for any final condition $P_f=(x_f,y_f)$, with $0\leq x_f^2+y_f^2\leq 1$, we denote by
 $t_{P_f}=t_{P_f}({\gamma})$ the optimal time to reach the fixed final condition, and by
$\omega_{P_f}=\omega_{P_f}({\gamma})$   the corresponding frequency $\omega$ of the optimal control (cf. (\ref{NSE}))

Using (\ref{curvex}) (\ref{curvey}), we know that
\be{curva}
\begin{array}{ccl}
x_f&= &x_{\omega_{P_f}({\gamma})}(t_{P_f}({\gamma}))  \\
y_f&= &y_{\omega_{P_f}({\gamma})}(t_{P_f}({\gamma}))
\end{array} \ee
as in equations (\ref{curvex}), (\ref{curvey}). And we let
$b_{P_f}({\gamma}):=1-\omega_{P_f}({\gamma})$ and $a_{P_f}({\gamma})=\sqrt{(b_{P_f}({\gamma}))^2+\gamma^2}$.

The function $t_{P_f}(\gamma)$ is monotonic non increasing since all controls available for $\gamma=\gamma_1$ are also available for any $\gamma \geq \gamma_1$.



We consider first the final condition {\it in the interior} of  the unit circle, so let $P_f=(x_f,y_f)$ be a fixed point such that $0\leq x_f^2+y_f^2<1$. We give  a bound on the value of the optimal frequency $\omega$ to be used in (\ref{NSE}). Define
$K_{P_f}:=\sqrt{\frac{x_f^2+y_f^2}{1-(x_f^2+y_f^2)}}$.

\bp{ovvia}
If $\omega\in \RR$ and $t>0$ is such that  $x_{\omega}(t)=x_f$ and $y_{\omega}(t)=y_f$, then
\be{ovviae}
1-\gamma K_{P_f} \leq \omega \leq 1+\gamma K_{P_f}.
\ee
\ep
\bpr
If (\ref{ovviae}) does not hold, then $b^2=(1-\omega)^2>\gamma^2K_{P_f}^2$. This implies (using $a^2:=\gamma^2+b^2$)
\[
\frac{\gamma^2}{a^2}\sin^2(at)<\frac{1}{1+K_{P_f}^2}.
\]
By using equations (\ref{curvex}) and (\ref{curvey}), we have:
\[
x_f^2+y_f^2=1-\frac{\gamma^2}{a^2}\sin^2(at)> \frac{K_{P_f}^2}{1+K_{P_f}^2}=x_f^2+y_f^2,
\]
which is a contradiction.
\epr
The above simple proposition gives bounds on the frequencies to be used for a given desired final condition $P_f$. This can be used in the search of the optimal control described in the previous section. In particular, if $P_f=(0,0)$, which corresponds to SWAP-like final conditions, $X_f:=\begin{pmatrix}0 & e^{i\phi}\cr e^{-i\phi} & 0 \end{pmatrix}$, then $K_{P_f}=0$, so the only admissible $\omega$ is $\omega=\omega_{P_f}(\gamma)=1$, independently of $\gamma$ (these are the resonant controls considered for example in \cite{OptRes}).

We now study the limit of $t_{P_f}(\gamma)$ as $\gamma$ goes to zero.
\bp{seconda}
If $P_f$ is in the interior of the unit disc, then
\be{t-gamma}
\lim_{\gamma \to 0^+}  t_{P_f}({\gamma})=+\infty.
\ee
If $P_f$ is on the boundary of the unit disc and corresponds to a phase $\psi_f$, i.e., $P_f=e^{i \psi_f}$, then
\be{T-circonferenza}
\lim_{\gamma \to 0^+}  t_{P_f}({\gamma})=\psi_f.
\ee
\ep

\bpr

Consider first the case where $P_f$ is inside the unit circle, i.e., the case of (\ref{t-gamma}).
Using equations (\ref{curvex}) and (\ref{curvey}),
we have that:
\be{relazione}
\frac{\gamma^2}{(b_{P_f})^2+\gamma^2}\sin^2\left(a_{P_f}\,t_{P_f}\right)
=1- x_f^2+y_f^2>0.
\ee

First we prove that
\be{omega-gamma}
\lim_{\gamma \to 0^+} |b_{P_f}({\gamma})|=0.
\ee

Assume, by the way of contradiction, that (\ref{omega-gamma}) does not hold. Then we have:
\[
\exists \,\epsilon >0,  \text{ such that }\forall \ n>0 \ \exists \ 0<\gamma_n<\frac{1}{n} \text{ with }
(b_{P_f}({\gamma_n}))^2 >\epsilon.
\]
This implies
\[
0\leq \frac{\gamma_n^2}{(b_{P_f}^{\gamma_n})^2+\gamma_n^2}\leq \frac{\gamma_n^2}{\epsilon+\gamma_n^2}
\]
Thus we have:
\[
\lim_{n\to+\infty} \frac{\gamma_n^2}{(b_{P_f}({\gamma_n}))^2+\gamma_n^2}
\sin^2\left(\sqrt{((b_{P_f}({\gamma_n}))^2+\gamma_n^2)}\,t_{P_f}(\gamma_n) \right)=0
\]
since the $\sin$ function is bounded, and $\gamma_n \to 0$. This contradicts equation (\ref{relazione}),
thus (\ref{omega-gamma}) holds.

Now we use (\ref{omega-gamma}) to prove (\ref{t-gamma}).
Assume, by the way of contradiction, that (\ref{t-gamma}) does not hold. As done before, this would imply that, for some $K>0$ there exists a sequence $\gamma_j$  ($j>0$) with $0<\gamma_j<\frac{1}{j}$, such that $0<t_{P_f}({\gamma_j})<K$.

Thus using (\ref{omega-gamma}):
\[
\lim_{j\to +\infty} \sqrt{(b_{P_f}({\gamma_j}))^2+\gamma_j^2}\,t_{P_f}({\gamma_j}) =0.
\]
Since $\frac{\gamma^2}{(b_{P_f}^{\gamma})^2+\gamma^2}<1$, the previous equation would imply that
\[
\lim_{j\to+\infty} \frac{\gamma_j^2}{(b_{P_f}({\gamma_j}))^2+\gamma_j^2}
\sin^2\left(\sqrt{(b_{P_f}({\gamma_j}))^2+\gamma_j^2}\,t_{P_f}({\gamma_j}) \right)=0
\]
and this, again, contradicts equation (\ref{relazione}).
Thus (\ref{t-gamma}) holds.

Finally for points on the unit circle (\ref{T-circonferenza}) follows from the explicit expression of the minimum time for a point $P_f=e^{i\psi_f}$ which was
derived in \cite{Noi} under the assumption $\gamma \leq 1$. This is
\be{T-circonferenza-bis}
t_{P_f}(\gamma)=\frac{\psi_f (2 \pi - \psi_f)}{\pi - \psi_f +\sqrt{\pi^2 + \gamma^2 \psi_f (2\pi - \psi_f)}}.
\ee
Taking the limit as $\gamma \rightarrow 0$ we obtain (\ref{T-circonferenza}).
\epr

\vs

In order to study the continuity of the function $t_{P_f}(\gamma)$ as a function of $\gamma$ (or as a function of $P_f$) we study, more generally, the continuity of the `joint' function $T(P_f, \gamma):=T(x_f,y_f,\gamma):=t_{P_f}(\gamma)$. This is a function defined on a half infinite cylinder ${\cal C}:=\{ x_f, y_f, \gamma \, | \, 0 \leq x_f^2+y_f^2 \leq 1, \, \gamma >0\}$. We shall restrict ourselves to study this function in the interior of ${\cal C}$, i.e., ${\texttt{int} \cal C}:= \{ x_f, y_f, \gamma \, | \, 0 \leq x_f^2+y_f^2 <  1, \, \gamma >0\}$.

\vs

\bp{continuita}
The function $T:=T(x_f,y_f,\gamma)$ is continuous at all points in $(P_f, \gamma) \in \texttt{int} {\cal C}$ such that $P_f$ is not on the critical trajectory corresponding to $\gamma$.
\ep

Recall from the results of \cite{Noi}, \cite{Raf} reviewed in the previous section that the critical trajectory is given by (\ref{curvex}) (\ref{curvey}) with $\omega:=\omega_c=1+\gamma^2$, and $0 \leq t \leq \frac{\pi}{2 a_c}:=\frac{\pi}{2 \gamma \sqrt{1+\gamma^2}}$. The locus of points of discontinuity is a surface in ${\cal C}$ whose intersection with the planes $\gamma=\texttt{const}$ is the critical trajectory at that $\gamma$, which is a spiral-like curve, which is very short for large $\gamma$ and very long for small $\gamma$.

\bpr
 For a given  $P_f$ and $\bar{\gamma}$, with $P_f$ not in the critical trajectory of $\bar{\gamma}$,
consider the corresponding values  $t_{P_f}(\bar{\gamma})$ and $\omega_{P_f}(\bar{\gamma})$ which are optimal. Then the two equations
\be{zero}
\begin{array}{llll}
F_1(t,\omega,x,y,\gamma):=x(t,\omega,\gamma)-x=0,\\
F_2(t,\omega,x,y,\gamma):=y(t,\omega,\gamma)-y=0,
 \end{array}
 \ee
hold at $t_{P_f}(\bar{\gamma}),\omega_{P_f}(\bar{\gamma}), x_f,y_f,\bar{\gamma}$. Moreover,
from the Implicit Mapping Theorem, they define in an open neighborhood $N$ of $(x_f,y_f,\bar \gamma)$ two continuous functions $t:=t(x,y,{\gamma})$, $\omega=\omega(x,y,\gamma)$, as long as the Jacobian with respect to the variables $t$ and $\omega$ is different from zero. We calculate using (\ref{curvex}) (\ref{curvey}),
\[
\text{Det } \left( \begin{array}{cc}
  \frac{\partial}{\partial t}F_1 & \frac{\partial}{\partial t}F_2 \\
  & \\

  \frac{\partial}{\partial \omega}F_1 & \frac{\partial}{\partial \omega}F_2
  \end{array} \right) = \frac{\gamma^2(\gamma^2+1-\omega)}{a^4} \sin( a t) \left[\sin( at)-at \cos (at)\right]
  \]
This expression is zero if and only if $t=0$, $t=\pi/a$ or $\omega=1+\gamma^2$.
Since the optimal time $t_{P_f}({\bar{\gamma}})$ is always positive and strictly less then $\frac{\pi}{a}$,\footnote{This follows from the results of \cite{Noi} \cite{Raf}. At $t=\frac{\pi}{a}$ every optimal candidate  trajectory is at the boundary of the unit circle where it looses optimality (if it had not lost it before by intersecting the critical trajectory).} and  the point $P_f$ does not belong to the critical trajectory, i.e., $\omega_{P_f}(\bar{\gamma})\neq {1+\bar{\gamma}^2}$, the Jacobian is not zero, so the two continuous functions are defined.
We know that
 $t(x_f,y_f,\bar \gamma)=t_{P_f}(\bar \gamma)$ and $\omega(x_f,y_f,\bar \gamma)=\omega_{P_f}(\bar \gamma)$. Let $V=\left(t(N),\omega(N)\right)$, this is a neighborhood of $\left(t_{P_f}(\bar \gamma), \omega_{P_f}(\bar \gamma)\right)$, moreover if $(t,\omega,x,y,\gamma)\in V \times N$, and satisfies the equations in (\ref{zero}), then necessarily $t=t(x,y, \gamma)$ and $\omega=\omega(x,y, \gamma)$.

  To show continuity  of the map $T$ at $(P_f,{\bar{\gamma}})$,i.e., of the time optimal function,
  we prove that there exists a
  neighborhood $W\subseteq N$, such that   if
 $ (x,y,\gamma)\in W$ then $T(x,y,\gamma)$ coincides with the implicit map $t(x,y,{\gamma})$, whose continuity is guaranteed by the Implicit Mapping  Theorem.

 It is obvious by definition that $t(x,y,{\gamma})\geq T(x,y,\gamma)$. Assume, by the way of contradiction, that
 the statement is false, then for all $n$ sufficiently large, there exists a sequence $\{P_n=(x_n,y_n)\}$ and a sequence $\{\gamma_n\}$, such that
 the sequence $P_n$ goes to $P_f$,  the sequence $\gamma_n$ goes to $\bar{\gamma}$, and $t(x_n,y_n,\gamma_n)>T(x_n,y_n,\gamma_n)$. On the other hand, we have:
 \[
 x_n=x\left(t(x_n,y_n,\gamma_n), \omega(x_n,y_n,\gamma_n),\gamma_n\right)=
 x\left(T(x_n,y_n,\gamma_n), \omega_{(x_n,y_n)}(\gamma_n),\gamma_n\right),
 \]
 \[
 y_n=y\left(t(x_n,y_n,\gamma_n), \omega(x_n,y_n,\gamma_n),\gamma_n\right)=
 y\left(T(x_n,y_n,\gamma_n), \omega_{(x_n,y_n)}(\gamma_n),\gamma_n\right),
 \]
where $\omega_{(x_n,y_n)}(\gamma_n)$ are the optimal values. Since $T(x_n,y_n,\gamma_n)$ and $\omega_{(x_n,y_n)}(\gamma_n)$ belong to compact sets (for the values $\omega$ see Proposition \ref{ovvia}), we may assume, after passing if necessary to a subsequence, that:
\[
\omega_{(x_n,y_n)}(\gamma_n) \to \bar{\omega},
\ \ T(x_n,y_n,\gamma_n) \to \bar{t}.\]
Since by continuity
\[
 x\left(T(x_n,y_n,\gamma_n), \omega_{(x_n,y_n)}(\gamma_n),\gamma_n\right) \to x(\bar{t},\bar{\omega},\bar{\gamma}), \  \   y\left(T(x_n,y_n,\gamma_n), \omega_{(x_n,y_n)}(\gamma_n),\gamma_n\right) \to y(\bar{t},\bar{\omega},\bar{\gamma}),
 \]
 we must have
\[
x_f=  x(\bar{t},\bar{\omega},\bar{\gamma}), \  \  y_f=  y(\bar{t},\bar{\omega},\bar{\gamma}).
\]
From the fact that  $t(x_n,y_n,\gamma_n)>T(x_n,y_n,\gamma_n)$, we have  $t_{P_f}(\bar \gamma)=t(x_f,y_f,\bar \gamma)\geq \bar{t}$, since $t_{P_f}(\bar \gamma)$ is optimal we conclude
$t_{P_f}(\bar \gamma)=t(x_f,y_f,\bar \gamma)= \bar{t}$. This, in turn, implies that also
$\bar{\omega}=\omega_{P_f}(\bar \gamma)$, since the optimal values is unique.\footnote{see \cite{Noi} or alternatively the geometric analysis of next section which shows that the value $\omega$ is the value of the parameter at the intersection of two parametric curves which is uniquely defined.}
Thus  for $n$ sufficiently large $\left(T(x_n,y_n,\gamma_n), \omega_{(x_n,y_n)}(\gamma_n)\right)$  belongs to $V$, thus we must have $T(x_n,y_n,\gamma_n)=t(x_n,y_n,\gamma_n)$ since the
 Implicit Map Theorem guarantees  uniqueness of  the function $t$, which contradicts $t(x_n,y_n,\gamma_n)>T(x_n,y_n,\gamma_n)$.

\epr


We now are interested in studying the discontinuity of the function $t_{P_f}=t_{P_f}(\gamma)$ at the points on the critical trajectory. The following result summarizes the continuity properties of this function.

\bp{discontin}
The function  $t_{P_f}=t_{P_f}(\gamma)$ is continuous for every $\gamma$ except for the $\gamma$'s such that $P_f$ is in the interior of  the critical trajectory. On these points it presents a discontinuity on the left and it is right continuous.
\ep

\bpr

First we first prove right continuity everywhere.

By monotonicity of the function $t_{P_f}$,  for a fixed value $\bar{\gamma}$, we know that:
\be{limitedestro}
\lim_{\gamma\to\bar{\gamma}^+} t_{P_f}({\gamma})=\sup_{\gamma>\bar{\gamma}}t_{P_f}({\gamma})=
l_+({\bar{\gamma}})\leq t_{P_f}({\bar{\gamma}}),
\ee
We will prove  that $l_+({\bar{\gamma}})= t_{P_f}({\bar{\gamma}})$.

By definition of $l_+({\bar{\gamma}})$, we know that for each $n\geq 1$ there exists $\bar{\gamma}<\gamma_n<\bar{\gamma}+\frac{1}{n}$, such that $l_+({\bar{\gamma}})-\frac{1}{n}<t_{P_f}({{\gamma_n}})<l_+({\bar{\gamma}})$.
Then for each $n\geq 1$ consider the corresponding optimal control value $\omega_{P_f}({{\gamma_n}})$.
By Proposition \ref{ovvia}, these control values are bounded, thus they admit a converging subsequence $\omega_{P_f}({{\gamma_{n_k}}})$. Along this subsequence we have that
$t_{P_f}({{\gamma_{n_k}}})\to l_+({\bar{\gamma}})$, $\omega_{P_f}({{\gamma_{n_k}}})\to \bar{\omega}$, and $\gamma_{n_k}\to \bar{\gamma}$.  By
using equations (\ref{curvex}) and (\ref{curvey}) we have
\[
x_f=x(\bar{\omega}, l_+^{\bar{\gamma}}) \   \  \text{ and } \  \ y_f=y(\bar{\omega},  l_+^{\bar{\gamma}}).
\]
Thus $ l_+({\bar{\gamma}})=t_{P_f}({\bar{\gamma}})$, otherwise $t_{P_f}({\bar{\gamma}})$ would not be the optimal time, since we have found a control that reaches $P_f=(x_f,y_f)$ in time $l_+({\bar{\gamma}})$.

Continuity of the function $t_{P_f}=t_{P_f}(\gamma)$,  for every $\gamma$ except for the $\gamma$'s such that $P_f$ is on the critical trajectory, follows from Proposition \ref{continuita}, where it is proved that the function is indeed continuous as a function of both variables $P_f$ and $\gamma$.

Assume now that $P_f$ is a point such that for $\gamma=\bar{\gamma}$ it lies in the interior of the critical trajectory. Therefore $t_{P_f}(\bar{\gamma})=\frac{ \alpha \pi}{2\bar{\gamma}\sqrt{(1+\bar{\gamma}^2)}}$ with $0< \alpha<1$. However, from the analysis in \cite{Noi}, \cite{Raf}, it follows that, for $\epsilon > 0$ sufficiently small, $t_{P_f}(\bar \gamma -\epsilon)> \frac{\pi}{2 (\bar \gamma - \epsilon) \sqrt{1+(\bar \gamma -\epsilon)^2}}$. This gives
 \be{discontinuita77}
 \lim_{\gamma \to \bar \gamma^-} t_{P_f}(\gamma)=\lim_{\epsilon \to 0+}  t_{P_f}(\bar \gamma -\epsilon) \geq
 \frac{\pi}{2 \bar \gamma \sqrt{1+ \bar \gamma^2}} >  \alpha \frac{\pi}{2 \bar \gamma \sqrt{1+ \bar \gamma^2}}. \ee
The analysis of \cite{Noi} \cite{Raf} also shows left continuity of $t_{P_f}$ if $P_f$ is exactly at the endpoint of the critical trajectory.\footnote{For $\gamma$ smaller than $\bar \gamma$ optimal trajectories reaching $P_f$ travel around the end point of the critical trajectory corresponding to that $\gamma$ before reaching $P_f$. The time to reach $P_f$ is therefore greater than the maximum time on the critical trajectory. However as $\gamma \to \bar \gamma^-$ the two points and the two times coincide.}

\epr

\vs

\br{Swapetal}
If $P_f$ corresponds to a SWAP like operator, then, since $P_f:=(0,0)$ is not on any critical trajectory,  $t_{P_f}$ is a continuous function of $\gamma$. For other points $P_f$ however in the interior of the unit disc, there are infinitely many values of $\gamma$ such that $P_f$ is on the corresponding critical trajectory.\footnote{They have a limit point at zero} At each of these values,  the function $t_{P_f}$ has a jump on the left. If we are trying to reach a point at exactly a time $T$ which is possibly  larger than the minimum time corresponding to a bound $\bar \gamma$, we cannot always  use the time optimal control for a $\gamma$ smaller than $\bar \gamma$. The time $T$ might not be in the range of the function $t_{P_f}$ even though such a function tends to $+\infty$ as $\gamma \rightarrow 0$ as shown in proposition \ref{seconda}. A characterization of the reachable sets for every $T$ will allow us to know exactly at what times we can reach a given state and how.
\er

\section{Geometry of the reachable sets} \label{Reachsets}

We now give a description of the reachable sets for system (\ref{basicmodel}), which will then be used in the solution of the minimum time  synchronization problem in the next section \ref{synchro}.  As the method has more general validity, we shall first describe it for general, bilinear,  right invariant systems on Lie groups and then specialize to system (\ref{basicmodel}).\footnote{Some of the concepts and ideas we shall describe are valid for more general families of vector fields. Restricting ourselves to bilinear right invariant vector fields ensure us that the solution of the associated initial value problems exists for every time $t$. For concreteness, our notation refers to {\it matrix} Lie groups, although we could have extended the discussion to abstract general Lie group simply replacing the exponential of a matrix with the exponential map.}

\subsection{General method}

Consider a system
\be{gensys2}
\dot X=AX+\sum_{j=1}^m u_jB_j X, \qquad X(0)=X_0,
\ee
where $A,B_1,\ldots,B_m$ are matrices in a matrix Lie algebra ${\cal L}$, $X$ belongs to the corresponding Lie group $e^{\cal L}$, $X_0$ is the given initial condition and $u_j$, $j=1,\ldots,m$  are the controls, which are assumed to belong to a set ${\cal U}$ of functions of time. The {\bf reachable set} at time $T$, ${\cal R}(T)$ is the set of states $X_f$ in $e^{\cal L}$ such that there exist functions $u_1,\ldots,u_m$ in ${\cal U}$ defined in $[0,T]$ so that the solution $X$ of  (\ref{gensys2}) with these controls satisfies  $X(T)=X_f$. ${\cal R}(0)=\{ X_0\}$  by definition and the reachable set ${\cal R}(\leq T)$ is defined as
\be{reaches}
{\cal R}(\leq T):=\bigcup_{0 \leq t \leq T} {\cal R}(t).
\ee
It follows obviously from the definition that the reachable sets ${\cal R}(\leq T)$ are non decreasing with $T$, i.e., ${\cal R}(\leq T_1) \subseteq {\cal R}(\leq T_2)$ if $T_1 \leq T_2$, a property not necessarily true for the reachable sets ${\cal R}(T)$.
\vs
To study reachable sets we can make a, possibly time varying, change of variables  in system (\ref{gensys2}) and study the reachable sets for the resulting system. The reachable sets for the original system are obtained by mapping back the reachable sets for the new system. In the case of system (\ref{gensys2}) we define\footnote{This is called `{\it passage to the interaction picture}' in the physics literature.}
\be{intepic}
U(t):=e^{-At}X(t).
\ee From (\ref{gensys2}), we obtain the differential equation for $U$,
\be{gensys3}
\dot U=\left(\sum_{j=1}^m u_j e^{-At} B_j e^{At} \right) U, \qquad U(0)=X_0.
\ee
Let ${\cal I}$ the smallest subspace of ${\cal L}$ which contains $\{B_1,\ldots,B_m\}$ and is invariant under Lie bracket with $A$, and let $\{ C_1,C_2,\ldots, C_l \}$ be a basis of ${\cal I}$. Then there are analytic functions $\gamma_{j,i}:=\gamma_{j,i}(t)$ such that
\be{llo2}
e^{-At} B_j e^{At}:=\sum_{i=1}^l \gamma_{j,i}(t)C_i.
\ee
By replacing this in (\ref{gensys3}) and defining
\be{Vi}
v_i:=\sum_{j=1}^m u_j \gamma_{j,i},
\ee
we obtain
\be{L3e}
\dot U=\left( \sum_{i=1}^l v_i C_i \right) U, \qquad U(0)=X_0.
\ee
If ${\cal V}$ denotes the image of the set of control functions, ${\cal U}$, under the map
(\ref{Vi}), then we can study the reachable sets for (\ref{L3e}) under the set of controls ${\cal V}$, denote them by ${\cal R}_U( T)$ and ${\cal R}_U(\leq T)$, respectively, and the reachable
sets for the original system (\ref{gensys2}), are recovered from (cf. (\ref{intepic}))
\be{recovering}
{\cal R}(T)=e^{AT} {\cal R}_U(T),\qquad {\cal R}(\leq T)=\bigcup_{0 \leq t \leq T} e^{At} {\cal R}_U(t).
\ee
This method is particularly useful when the set of controls ${\cal V}$ has the {\it scalability property}, that is, $\vec v \in {\cal V}$ implies $L_\alpha({\vec v}):= \alpha \vec v (\alpha t) \in {\cal V}$ for every $\alpha \in [0,1]$. In this case, the reachable sets ${\cal R}_U(T)$ are increasing with $T$, so that ${\cal R}_U(T)={\cal R}_U(\leq T)$ for every $T$. This is seen with a standard argument for driftless systems such as (\ref{L3e}). Let $\vec v$ a control in ${\cal V}$ steering the initial condition $X_0$ to $U_f$ in time $T_1$, so that $U_f \in {\cal R}_U(T_1)$. Let $T_2 > T_1$ and consider the control $L_{\frac{T_1}{T_2}} \vec v(t)=\frac{T_1}{T_2} \vec v(\frac{T_1}{T_2}t)$. With this control the function $\tilde U(t):=U(\frac{T_1}{T_2}t)$ solves (\ref{L3e}) and it is such that $\tilde U(T_2)=U_f$, so that $U_f \in {\cal R}_U(T_2)$.

In general, studying ${\cal R}(\leq T)$ is easier than studying ${\cal R}(T)$ because this set is related to the solution of the time optimal control problem. It is a well known fact in geometric control theory that if $X_f$ is a final point of a time optimal trajectory at time $T$, then $X_f$ is on the boundary of the reachable ${\cal R}(\leq T)$. On the other hand, by definition, the minimum time to reach $X_f$, is the smallest time $T$ such that $X_f \in {\cal R}(T)$. In the above described situation, the minimum time $T$ is the smallest time $t$ such that
\be{L4r}
X_f \in {\cal R}(t)=e^{At} {\cal R}_U(t),
\ee
and, if we have a description of ${\cal R}_U(t)(={\cal R}_U(\leq t))$ this gives an alternative  way to find the minimum time and control.

\subsection{Reachable sets for systems on $SU(2)$}

We now apply the strategy outlined above to the case of system (\ref{basicmodel}). In this case,  the space of controls, ${\cal U}$ is the space of Lebesgue measurable functions $\vec u:=(u_x,u_y)$ with Euclidean norm $\| \vec u \|:=\sqrt{u^2_x+  u^2_y} \leq \gamma$. Specializing (\ref{llo2}), we obtain
\be{L1X}
e^{-\tilde \sigma_z \tau} \tilde \sigma_x e^{\tilde \sigma_z \tau}=\cos(\tau) \tilde \sigma_x -\sin(\tau)\tilde \sigma_y,
\ee
and
\be{L1Y}
e^{-\tilde \sigma_z \tau} \tilde \sigma_y e^{\tilde \sigma_z \tau}=\sin(\tau) \tilde \sigma_x+ \cos(\tau) \tilde \sigma_y.
\ee
Therefore, the equation corresponding to (\ref{L3e}) is
\be{L5bis}
\dot U=(v_x \tilde \sigma_x + v_y \sigma_y )U, \qquad U(0)= {\bf 1}
\ee
with $v_x:=\cos(\tau) u_x+ \sin(\tau) u_y$ and $v_y=-\sin(\tau) u_x + \cos(\tau) u_y$. Therefore, if $\vec u:=[u_x,u_y]^T$ and $\vec v:=[v_x,v_y]^T$, we have
\be{relat32}
\vec v=\begin{pmatrix}  \cos(\tau) & \sin(\tau) \cr -\sin(\tau) & \cos(\tau) \end{pmatrix} \vec u.
\ee
 Formula (\ref{relat32}) gives a one to one correspondence between Lebesgue measurable functions $\vec u$ with norm bounded by $\gamma$ and  Lebesgue measurable functions $\vec v$ with norm bounded by $\gamma$. That is, in this case, the space of controls ${\cal V}$ coincides with ${\cal U}$. Moreover, ${\cal V}$ has the scalability property, and therefore ${\cal R}_U(\leq T)={\cal R}_U(T)$ for every $T\geq 0$.

 The following theorem describes ${\cal R}_U(T)$ for system (\ref{L5bis}), from which, in the following corollary, we obtain the reachable sets for the original system (\ref{basicmodel}). As we have already done in the previous section, we scale the time variable as $t:=\frac{\tau}{2}$ and we we refer to `time' (as for example for $T$ in ${\cal R}_U(T)$) we shall refer to the time $t$ defined this way.

\begin{figure}[htb]
\centering
\includegraphics[width=0.7\textwidth, height=0.45\textheight]{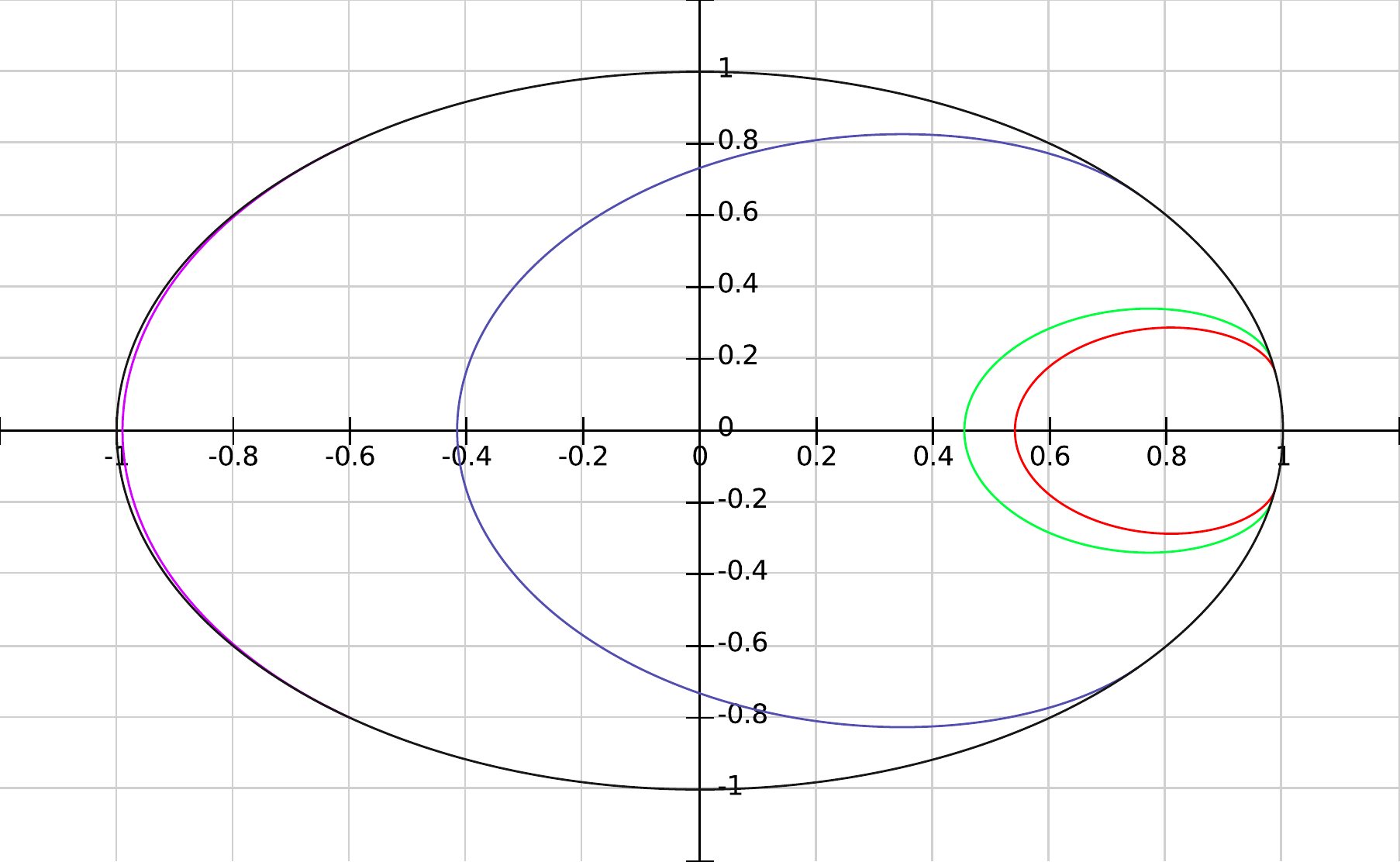}
\caption{Boundaries of reachable sets ${\cal R}_U(\leq T)={\cal R}_U(T)$ for various values of $T$. The parameter $\gamma$ is chosen equal to $1$. In red, we have the boundary (curves (\ref{CurveX9}), (\ref{CurveY9})) for $T=1$, in green for $T=1.1$, in blue for $T=2$, in purple for $T=3$. For $T=\frac{\pi}{\gamma}=\pi$ the boundary of the reachable set collapses to the single point $(-1,0)$ and the reachable set becomes the whole unit disc and therefore the whole $SU(2)$. }
\label{Fig2}
\end{figure}

\bt{Reachset1} The reachable set ${\cal R}_U(T)$ for the system (\ref{L5bis}) is given by the set of matrices in $SU(2)$ with the $x_{1,1}:=x+iy$ entry in the region of the unit disc bounded by the parametric curve ${\cal F}_T$ defined as.\footnote{These are the optimal frontlines studies in \cite{Raf}.}
\be{CurveX9}
x=x(\omega)=\cos(\omega T) \cos(aT)+\frac{\omega}{a} \sin(\omega T) \sin(aT),
\ee
\be{CurveY9}
y=y(\omega)=\sin(\omega T) \cos(aT)-\frac{\omega}{a} \cos(\omega T) \sin(aT),
\ee
where $a:=\sqrt{\omega^2+\gamma^2}$ and the parameter $\omega \in \left[ -\sqrt{\frac{\pi^2}{T^2}-\gamma^2}, \sqrt{\frac{\pi^2}{T^2}-\gamma^2} \right]$.
\et
The region of the unit disc representing ${\cal R}_U(T)$ is at the right of the curve ${\cal F}_T$ which, for every $T$, connects two points on the boundary of the unit disc and is symmetric with respect to the $x$-axis. The region grows with $T$, and when $T=\frac{\pi}{\gamma}$, the curve ${\cal F}_T$ collapses to the point $(-1,0)$ and the reachable set ${\cal R}_U(T)$ becomes all of $SU(2)$. Figure \ref{Fig2} shows the typical behavior for the curves ${\cal F}_T$ for various values of $T$.

Specializing (\ref{recovering}) we obtain
\bc{special}
\be{cf}
{\cal R}(T)=e^{2 \tilde \sigma_z T} {\cal R}_U(T).
\ee
In other terms, $X_f \in {\cal R}(T)$ if and only if, denoted by $P_f$ the $(1,1)$ entry of $X_f$, we have that    $e^{-iT} P_f$ is in the region described in Theorem \ref{Reachset1}.
\ec
\br{explanationdisc}
The above corollary gives an alternative geometric explanation of the discontinuity we described in the previous section (cf. Remark \ref{Swapetal}). Consider the desired point $P_f$, in the unit disc. As $T$ increases $e^{-iT} P_f$ describes  a circle inside the unit disc. At the same time, as $T$ increases the curve ${\cal F}_T$ moves towards left with a speed which decreases with decreasing $\gamma$. The optimal time is the minimum time where these two curves intersect for the first time. If this intersection happens at a point of tangency, a small decrease in $\gamma$ implies that $e^{-iT} P_f$ will have to go around almost an entire circle again before intersecting ${\cal F}_T$, which explains the discontinuity at that $\gamma$. Notice that this problem does not occur at the origin (corresponding to SWAP-like operators) since in this case the circle $e^{-iTP_f}$ reduces to a single point.
\er

To prove Theorem \ref{Reachset1} we shall need the following facts about the parametric curves ${\cal F}_T$ (with parameter $\omega$ and fixed $T$), and its extension to  values (in (\ref{CurveX9}), (\ref{CurveY9})) of ${\omega} \in (-\infty, -\sqrt{\frac{\pi^2}{T^2}-\gamma^2}) \bigcup (\sqrt{\frac{\pi^2}{T^2}-\gamma^2},  +\infty)$,  which we denote by ${\cal S}_T$.  It can be inferred by plotting  the curves for different values of $T$, but we present an analytic proof in the appendix.

\newpage

\bl{SIO}
Let ${\cal F}_t$  and ${\cal S}_t$ the previous  defined curves, then:
\begin{enumerate}
\item If $t_1\neq t_2$ then  ${\cal F}_{t_1}\cap {\cal F}_{t_2}=\emptyset$
\item The curve ${\cal F}_t$ does not have self intersections.
\item For any $0<t< \frac{\pi}{\gamma}$, we have ${\cal F}_{t} \cap {\cal S}_t=\emptyset$.
\end{enumerate}
\el

\bpr (Proof of Theorem \ref{Reachset1}) Consider the problem of time optimal control subject to $v_x^2+v_y^2 \leq \gamma^2$. Analogously to the case of the system with drift considered in \cite{Noi} \cite{Raf} and summarized in section \ref{Rev}, the candidate optimal controls have the form (cf. (\ref{NSE}))\footnote{No singular extremals of the form $v_x=v_y\equiv0$ are admissible in this case since they would mean zero dynamics and therefore they are incompatible with the minimum time requirement.}
\be{NSE1}
v_x=\gamma \sin(\omega \tau + \tilde \phi), \qquad v_y=-\gamma \cos(\omega \tau +\tilde \phi).
\ee
As in the case of system with drift of section \ref{Rev}, with this control, the differential equation (\ref{L5bis}) can be explicitly integrated. The result is (cf. (\ref{soluzexpli}))
\be{soluzexpli3}
U(\tau,\omega,\tilde \phi):=\begin{pmatrix} e^{i \omega t}(\cos(a t)- i \frac{\omega}{a} \sin(a t)) & e^{i (\omega t + \tilde \phi)} \frac{\gamma}{a} \sin(a t) \cr  - e^{-i(\omega t + \tilde \phi)} \frac{\gamma}{a} \sin(a t)  & e^{-i \omega t}(\cos(a t)+i \frac{\omega}{a} \sin(a t)) \end{pmatrix},
\ee
with $a:=\sqrt{\omega^2+\gamma^2}$.
From this expression, it follows that, if a final condition is reached optimally in time $T$, the same is true (by simply changing the phase $\tilde \phi$) for any state which differs from it only by the phase of the off diagonal element. Therefore, we only have to consider the (1,1) entry of the desired final condition. This is a point inside the unit disc with coordinates (cf. (\ref{curvex}), (\ref{curvey}) and (\ref{CurveX9}) (\ref{CurveY9}))
\be{coordi1}
x=x_{\omega}(t)=\cos(\omega t) \cos(at)+\frac{\omega}{a} \sin(\omega t) \sin(at),
\ee
\be{coordi2}
y=y_{\omega}(t)=\sin(\omega t) \cos(at)-\frac{\omega}{a} \cos(\omega t) \sin(at).
\ee

\vs


Consider now the curve (\ref{coordi1}) (\ref{coordi2}) corresponding to $t$, and with $\omega \in \left[ - \sqrt{\frac{\pi^2}{t^2}-\gamma^2}, \sqrt{\frac{\pi^2}{t^2}-\gamma^2} \right]$, which connects two symmetric points on the unit circle, i.e. ${\cal F}_t$.


From Lemma \ref{SIO}, we know that ${\cal F}_{t_1}$ and ${\cal F}_{t_2}$ do not intersect for $t_1 < t_2$. Moreover by comparing the $x$ values of the endpoints of ${\cal F}_{t_1}$ and ${\cal F}_{t_2}$ for $t_1$ and $t_2$, which are $x_1=-\cos(\sqrt{\pi^2 -\gamma^2 t_1^2})$ and $x_2=-\cos(\sqrt{\pi^2 -\gamma^2 t_2^2})$, it follows that $x_2 < x_1$. So, if $t_1 < t_2$ ${\cal F}_{t_1}$ lies entirely to the right of ${\cal F}_{t_2}$.   Now we want to show that ${\cal S}_{t_1}$, also is entirely to the right of ${\cal F}_{t_2}$. In fact, since $\lim_{\omega \to \pm \infty} x_\omega(t)=1$ and $\lim_{\omega \to \pm \infty} y_\omega(t)=0$,  ${\cal S}_{t_1}$ eventually belongs to the region on the right of  ${\cal F}_{t_1}$. If ${\cal S}_{t_1}$ where to intersect ${\cal F}_{t_2}$, then would have to `go back' to the region on the right of ${\cal F}_{t_1}$ and therefore intersect ${\cal F}_{t_1}$, something which is excluded by Lemma \ref{SIO}.

In view of this fact, for every $(x_0,y_0) \in {\cal F}_T$,  $T$ is the smallest time such that $(x_0, y_0)$ can be reached with an extremal control, and therefore it is the minimum time to reach $(x_0,y_0)$. As a consequence, $(x_0,y_0)$ is on the boundary of the reachable set ${\cal R}_U(\leq T)={\cal R}_U( T)$.\footnote{or, more properly, the projection of the reachable set onto the unit disc where the actual reachable set in $SU(2)$ is the set of all the matrices whose (1,1) entry is in the region bounded by the curve ${\cal F}_T$.}

\epr

Using Theorem \ref{Reachset1} we obtain an alternative (yet related) method to the ones in \cite{Noi}, \cite{Raf} to find the minimum time and the optimal control to reach a final condition $X_f$ for system (\ref{basicmodel}) represented by a point $P_f$ in the unit disc. The minimum time is the minimum $t$ such that
\be{formmintim}
e^{-it} P_f \in {{\cal  R}_U(t)}.
\ee

\vs

The optimal control problem with the method from formula (\ref{formmintim}) can be solved graphically or numerically by finding for increasing values of $T$ the distance from the point $e^{-iT}P_f$ and the curve ${\cal F}_t$ until such a distance becomes zero. The first time $T$ where this distance is zero is the minimum time. The value of $\omega$ of the intersection gives the frequency to be used in the controls, $v_x$ and $v_y$. Inverting the transformation (\ref{relat32}),  we obtain the control for the original system (\ref{basicmodel}), which is of the form (\ref{NSE}).\footnote{Notice that the frequency $\omega$ for the driftless system and the frequency $\omega$ for the original system with drift are not the same. We should have used two different symbols for $\omega$ (as well as for $a$) but we have not done it to keep the notations simple.} Finally the phase is $\tilde{\phi}$ is found in (\ref{NSE}) to match the desired final condition in $SU(2)$ according to (\ref{soluzexpli}).


\section{Solution of the minimum time synchronization problem for $N$ qubits}\label{synchro}

With the above description of the reachable set for system (\ref{basicmodel}) we can now solve the problem to drive simultaneously to the desired final conditions, $N$ qubits in minimum time. Let us assume the bounds on the norm of the controls $\sqrt{u_x^2+u_y^2}$ are $\gamma_1,\ldots,\gamma_N$ for system $1,\ldots,N$,\footnote{This takes into account possible differences in the Larmor frequencies of spin $\frac{1}{2}$ particles since the equations are normalized.}. Let $X_{f,1},\ldots,X_{f,N}$ be the desired  final conditions for system $1,\ldots,N$, respectively. We also denote by ${\cal R}_j(t)$ the reachable set at time $t$  for system $j$.

The first step to find the time optimal synchronous control is to solve independently the time optimal control problems for the systems $1$ through $N$, and to find therefore the minimum times $T_1$,...,$T_N$. This can be done using the methods of \cite{Noi}, \cite{Raf}, or the method based on the evolution of the reachable set in the previous section. Let us consider now the maximum among $T_1,...,T_N$; Assume it is $T_1$. Then for $j=2,\ldots,N$ we check whether $X_{f,j} \in {\cal R}_j(T_1)$. If that is the case, then the minimum time is $T_1$ and we can simply drive system $1$, time optimally and (possibly) slow down systems $2$ through $N$ in order to obtain the desired final conditions $X_{f,2}$,...,$X_{f,N}$ in time $T_1$. To slow down a system, we can apply the optimal control but with a smaller value of $\gamma$. Graphically $e^{-i T} P_f$ is in the region bounded by ${\cal F}_T$. If we consider a smaller $\gamma$ the boundary of this region is moved towards the right. We can do that, until the boundary includes the point $e^{-iT}P_f$ and this gives us the correct $\gamma$ to use.

Assume instead that there exists a $\bar j \in \{2,\ldots,N\}$ such that $X_{f,\bar j}\notin {\cal R}_{\bar j}(T_1)$. Then the minimum time for synchronous control is greater than $T_1$ and greater than or equal to the smallest $T> T_1$ such that $X_{f,\bar j} \in {\cal R}_{\bar j}(T)$. Find such a $T$ and then check that for all other $j=1,\ldots,N$, $j \not= \bar j$, $X_{f,j} \in {\cal R}_j(T)$. If that is the case, the algorithm stops otherwise it continues going back and replacing $T_1$ with $T$. The algorithms ends because for $T \geq \frac{\pi}{\gamma_{min}}$, where $\gamma_{min}:=\min\{\gamma_1,\ldots,\gamma_N\}$, ${\cal R}_j(T)=SU(2)$, for every $j$.

\vs

We summarize the procedure in the following formal algorithm.

\vs

{\scshape

\vs

{\bf ALGORITHM}

\vs
\begin{enumerate}

\item Solve the Time Optimal Control Problem for systems 1 through N and find the minimum
times $T_1,\ldots,T_N$.

Set $T_{curr}:=\max \{T_1,\ldots,T_N\}$.

Set $k_{curr}$ a value of $j$ such that $T_j=T_{curr}$.

\item  For $j \not=k_{curr}$ check that

  $X_{f,j}  \in {\cal R}_j(T_{curr})$

  If this is the case then STOP.

\item Choose a $\bar j \in \{1,\ldots,N\}$  such that

  $X_{f,\bar j}  \notin {\cal R}_{\bar j}(T_{curr})$

\item Find the smallest $T > T_{curr}$ such that

  $X_{f,\bar j}  \in {\cal R}_{\bar j}(T)$

\item Set $T_{curr}=T$, $k_{curr}=\bar j$

\item  Go back to step 2.

\end{enumerate}
}



\section*{Acknowledgement} Domenico D'Alessandro's  research was  supported by ARO MURI grant W911NF-11-1-0268. He also would like to thank Dr. R. Romano for useful discussions. The authors
 would like to thank Prof. B. Jakubczyk for helpful comments which went as far as providing meaningful examples for  some of the properties we wanted to prove in this paper.

\section*{Appendix: Proof of Lemma \ref{SIO}}

First we rewrite, for convenience, the equations of the curves   ${\cal{F}}_t$ and ${\cal{S}}_t$.
We have:
\be{ap-coordi1}
x=x_{\omega}(t)=\cos(\omega t) \cos(at)+\frac{\omega}{a} \sin(\omega t) \sin(at),
\ee
\be{ap-coordi2}
y=y_{\omega}(t)=\sin(\omega t) \cos(at)-\frac{\omega}{a} \cos(\omega t) \sin(at).
\ee
Then ${\cal{F}}_t$ is the curve $\left(x_{\omega}(t),y_{\omega}(t)\right)$ for $\omega\in  \left[ - \sqrt{\frac{\pi^2}{t^2}-\gamma^2}, \sqrt{\frac{\pi^2}{t^2}-\gamma^2} \right]$, while ${\cal{S}}_t$ is (\ref{ap-coordi1}),  (\ref{ap-coordi2}), with $\omega \in (-\infty,  - \sqrt{\frac{\pi^2}{t^2}-\gamma^2}) \cup
(\sqrt{\frac{\pi^2}{t^2}-\gamma^2}, +\infty)$.

\subsection*{ 1. If $t_1\neq t_2$ then  ${\cal F}_{t_1}\cap {\cal F}_{t_2}=\emptyset$}

For $\omega=-\sqrt{\frac{\pi^2}{t^2}-\gamma^2}$ we have that:
\be{ap-INTPOIN1}
x=-\cos(\sqrt{\pi^2 - \gamma^2 t^2}) , \  \  \  y=\sin(\sqrt{\pi^2 - \gamma^2 t^2}),
\ee
which is a point on the unit circle. Since
\be{simmetria}
x_{-\omega}(t)=x_{\omega}(t), \  \   \  y_{-\omega}(t)=-y_{\omega}(t),
\ee
the curve   ${\cal{F}}_t$ connects two symmetric points on the unit circle.

Consider now two curves  ${\cal F}_{t_1}$ and ${\cal F}_{t_2}$, and let us assume $t_1 < t_2$.
To prove that these two curves do not have intersections, we  actually need to check that they do not have intersections  only for $\omega \in \left[ -\sqrt{\frac{\pi^2}{t^2}-\gamma^2}, 0 \right]$,
since they are symmetric with respect to the $x$ axis (see equation (\ref{simmetria})).
 First notice that at the endpoint, when $\omega=-\sqrt{\frac{\pi^2}{t^2}-\gamma^2}$, on the unit circle,
equation (\ref{ap-INTPOIN1}), implies that the $x$ coordinate is
{\it strictly decreasing with $t$}, therefore the intersection point cannot happen on the unit circle. For $\omega=0$, $y=0$  for all $t$, and  $x=\cos(\gamma t)$, which is also strictly decreasing with $t$. So, if an intersection point between ${\cal F}_{t_1}$ and ${\cal F}_{t_2}$ occurs it cannot be on the boundary of the unit disc or on the $x$ axis. It has to be for a parameter $\omega:=\omega_1$ for ${\cal F}_{t_1}$ in $\left(-\sqrt{\frac{\pi^2}{t_1^2}-\gamma^2}, 0 \right)$ and for a parameter $\omega:=\omega_2$ for ${\cal F}_{t_2}$ in $\left(-\sqrt{\frac{\pi^2}{t_2^2}-\gamma^2}, 0 \right)$.

To show that this is also not possible, we first show that it cannot exist a point where  ${\cal F}_{t_1}$ and ${\cal F}_{t_2}$ are tangent to each other. Calculation of $\frac{dx}{d\omega}$ gives
\be{dxdom}
\frac{dx}{d\omega}=\frac{\gamma^2}{a^3} \sin(\omega t) \left[-a t \cos(at)+\sin(at)
\right];
\ee
Calculation of $\frac{dy}{d\omega}$ gives
\be{dydom}
\frac{dy}{d\omega}=-\frac{\gamma^2}{a^3} \cos(\omega t) \left[-a t \cos(at)+\sin(at)
\right].
\ee
The function $f(at)=-a t \cos(at)+\sin(at)$ is always positive for $at \in [\gamma t, \pi]$ (recall we are assuming $0 < \gamma t <\pi$). Since $\omega$ is negative and $|\omega t|\leq  \sqrt{\pi^2-\gamma^2t^2}<\pi$,  $\frac{dx}{d\omega}<0$ and the curve ${\cal F}_t$ gives $y$ as a well defined function of $x$. Its derivative is
\be{deryx}
\frac{dy}{dx}=\frac{\frac{dy}{d\omega}}{\frac{dx}{d\omega}}=-\cot(\omega t).
\ee
This function is always increasing in the considered interval (it goes from $\cot(\sqrt{\pi^2 -\gamma^2 t})$ to $+\infty$). Therefore if the curves ${\cal F}_{t_1}$ and ${\cal F}_{t_2}$ coincide so that their slope are the same, we must have
\be{L1}
\omega_1 t_1=\omega_2 t_2.
\ee
Moreover the points where they intersect must coincide. Working in polar coordinates, the radius of a point from equations (\ref{ap-coordi1}) and (\ref{ap-coordi2}) is given by $r^2=1-\frac{\gamma^2}{a^2} \sin^2(at)$. Equalities of the radiuses gives ($a_{1,2}=\sqrt{\omega_{1,2}^2+\gamma^2}$).
\be{L2}
\frac{1}{a_1^2} \sin^2(a_1 t_1)=\frac{1}{a_2^2} \sin^2(a_2 t_2).
\ee
The phase of the point is given by (cf. (\ref{soluzexpli3}))
\be{fase}
\psi=\omega t+\delta-\arctan\left(\frac{\omega}{a}\tan(at)\right)+2 k \pi,
\ee
where $\delta=0$ or $\delta=\pi$ according to whether the point is in the first or second quadrant.\footnote{To be more precise, we should have replaced $\arctan\left(\frac{\omega}{a}\tan(at)\right)$ with $-\frac{\pi}{2}$ in the case where $at=\frac{\pi}{2}$. The treatment of this special case is analogous to the treatment of the case where $at\not= \frac{\pi}{2}$. So we focus on this case only.} Therefore we must have
\be{oop423}
\omega_1 t_1 +\delta - \arctan\left(\frac{\omega_1}{a_1} \tan(a_1 t_1)\right)+2 k_1 \pi=
\omega_2 t_2 +\delta - \arctan\left(\frac{\omega_2}{a_2} \tan(a_2 t_2)\right)+2 k_2 \pi.
\ee
Since $|\omega_{1,2} t_{1,2}| \leq \sqrt{\pi^2-\gamma^2t_{1,2}^2} < \pi$ the absolute value of the sum of the first and third terms in both left and right hand side formula (\ref{oop423}) is bounded by $\frac{3 \pi}{2}$. So the above equality (\ref{oop423}) can only occur for $k_1=k_2$. This implies
\be{oop423e}
\omega_1 t_1 - \arctan\left(\frac{\omega_1}{a_1} \tan(a_1 t_1)\right)=
\omega_2 t_2 - \arctan\left(\frac{\omega_2}{a_2} \tan(a_2 t_2)\right).
\ee
Using (\ref{L1}) and the fact that $\arctan$ is an increasing function, we obtain,
\be{L3}
\frac{\omega_1}{a_1} \tan(a_1 t_1)=\frac{\omega_2}{a_2} \tan(a_2 t_2).
\ee
Using (\ref{L2}) and the fact that $0< at< \pi$, we obtain from (\ref{L3})
\be{L4}
\frac{\cos^2(a_1 t_1)}{\omega_1^2}=\frac{\cos^2(a_2 t_2)}{\omega_2^2}.
\ee
Combining (\ref{L2}) and (\ref{L4}), we obtain
\be{L5}
1=\frac{\omega_1^2}{\omega_2^2} \cos^2(a_2 t_2)+ \frac{a_1^2}{a_2^2} \sin^2(a_2 t_2).
\ee
From (\ref{L1}) and $t_1 < t_2$ we obtain $\frac{\omega_1^2}{\omega_2^2} >1$ and $\frac{a_1^2}{a_2^2} >1$. Therefore in (\ref{L5}), we have
\be{L6}
1>\cos^2(a_2 t_2)+ \sin^2(a_2 t_2) =1,
\ee
which is a contradiction.

\vs

So, if an intersection point between ${\cal F}_{t_1}$ and ${\cal F}_{t_2}$ occurs, at the intersection the two curves are not tangent. Now we use this to prove that intersections are not possible.

Denote by $C_1=\{t\in (0,t_2)\ | \  {\cal F}_{t}\cap{\cal F}_{t_2}\neq \emptyset \}$, and by $C_2$ its complement.
We need to show that $C_1=\emptyset$. We will   prove that this set is both closed and open, so it must be empty since the interval $(0,t_2)$ is a connected set.

First we prove that $C_1$ is open.
Denote by $\cal{A}$ the region bounded by the unit circle, the $x-$axis, and the curve ${\cal F}_{t_2}$.
If a curve ${\cal F}_{\tilde{t}}$, for $0<\tilde{t}<t_2$ intersects the curve ${\cal F}_{t_2}$, since at the intersection point the two curves are not tangent, this implies that the curve ${\cal F}_{\tilde{t}}$ leaves the region $\cal{A}$.
Let $Q=\left(x_{{\tilde\omega}}(\tilde{t}),y_{{\tilde\omega}}(\tilde{t})\right)\in {{\cal{F}}_{\tilde{t}}}$ and outside $\cal A$.
Let $V$ be   a neighborhood   of $Q$ which lies outside ${\cal A}$.  By continuity there exits $\epsilon>0$ such that if $|t-\tilde{t}|\leq\epsilon$ and $|\omega-\tilde\omega|<\epsilon$, then $\left(x_{ {\omega }}(t ),y_{ {\omega}}(t )\right)\in V$.  By choosing, if necessary, $\epsilon_1<\epsilon$, such that $\omega\in  \left[ - \sqrt{\frac{\pi^2}{t^2}-\gamma^2},0\right]$, for $|t-\tilde{t}|\leq\epsilon_1$ and $|\omega-\tilde\omega|<\epsilon_1$, we have that for $t\in (\tilde{t}-\epsilon_1,\tilde{t}+\epsilon_1)$
the curve ${\cal{F}}_t$ reaches $V$, so goes outside ${\cal A}$, and so must intersect  ${\cal F}_{t_2}$.
Thus $C_1$ is open.

Now  we prove that also $C_2$ (the complement of $C_1$) is an open set. If $\tilde{t}\in C_2$, then the curve ${\cal{F}}_{\tilde{t}}$ lies all inside the interior of ${\cal A}$. Thus there exists a neighborhood $W$ of this curve which lies all inside ${\cal A}$.
For all $\tilde\omega\in  \left[ - \sqrt{\frac{\pi^2}{\tilde{t}^2}-\gamma^2},0\right]$ there exists $\delta>0$ such that if $|t-\tilde{t}|\leq\delta$, and $|\omega-\tilde\omega|<\delta$ then $(x_{\omega}(t),y_{\omega}(t))\in W$.
The constant $\delta$ depends on $\tilde{t}$ and also on $\tilde\omega$. Since $\tilde\omega$ varies in a compact set, we may choose a common $\delta>0$, thus all the curves ${\cal{F}}_t$ for $t\in (\tilde{t}-\delta,\tilde{t}+\delta)$ lie in $W$, so in particular they do not intersect ${\cal{F}}_{t_2}$. So $(\tilde{t}-\delta,\tilde{t}+\delta)\subset C_2$, so $C_2$ is open, and this
 implies that $C_1$ is closed.

\vs

\subsection*{2.  The curve ${\cal F}_t$ does not have self intersections.}

Consider the function $r^2=1-\frac{\gamma^2}{a^2}\sin^2(at)$ seen as a function of $\omega$. This function is even and  it is easily seen, by taking the derivative with respect to $\omega$, that this is decreasing in the interval $
[ - \sqrt{\frac{\pi^2}{t^2}-\gamma^2},0]$ (so consequently  increasing in the interval $[0,  \sqrt{\frac{\pi^2}{t^2}-\gamma^2}]$). Thus a possible self intersection may only be for two opposite values of $\omega$.
On the other hand, by equation (\ref{simmetria}), to have equality we must suppose $y_{\omega}(t)=0$ and this can happen only for $\omega=0$.\footnote{This is easily seen as follows: If $y=0$, from (\ref{ap-coordi2}) we have $\sin(\omega t) \cos(at)=\frac{\omega}{a} \cos(\omega t) \sin(at)$. From this, if $\cos(\omega t)=0$ then $\cos(at)=0$ and viceversa. Therefore in this case, we would have (take positive $\omega$) $\omega t=\frac{k\pi}{2}$ and $at=\frac{l\pi}{2}$ with $k$ and $l$ odd and strictly positive (since we are assuming $\omega\not=0$). Solving for $t$, we get $\frac{\omega}{a}=\frac{k}{l}$ and using $a=\sqrt{\omega^2+\gamma^2}$ we get $(l^2-k^2) \omega^2 =k^2 \gamma^2$ which shows that $l > k$. From this, we also obtain $\omega=\sqrt{\frac{k^2}{l^2 - k^2}}\gamma$ which replaced in $\omega t=k \frac{\pi}{2}$ gives $t=\frac{\pi \sqrt{l^2 -k^2}}{2 \gamma}$ that along with the fact that $l$ and $k$ are odd and $l$ is strictly greater than $k$ contradicts the fact that $t < \frac{\pi}{\gamma}$. Therefore, we can write $\frac{\tan(\omega t)}{\omega t}=\frac{\tan(a t)}{a t}$. Since for the given bounds on the value of $\omega t$ is $ 0 < \omega t < \sqrt{\pi^2 -\gamma^2t}< \pi$, and the value of $at$ is $ \gamma t < at < \pi$, so the equality $\frac{\tan(\omega t)}{\omega t}=\frac{\tan(a t)}{a t}$ means that $\omega t$ and $at$ are both in $(0,\frac{pi}{2})$ or both in $(\frac{\pi}{2}, \pi)$. On both these intervals the function $\frac{\tan(x)}{x}$ is increasing. Therefore we have $\omega t=at$ which is possible only if $\omega=0$ and $\gamma=0$.} So ${\cal{F}}_t$ does not have self intersections.

\subsection*{3. For any $0<t< \frac{\pi}{\gamma}$, we have ${\cal F}_{t} \cap {\cal S}_t=\emptyset$. }

We consider positive $\omega$ because of the symmetry of the curve (\ref{ap-coordi1}) (\ref{ap-coordi2}). We want to show that ${\cal S}_t$ remains strictly inside the region of the unit disc bounded by ${\cal F}_{{t}}$ and the boundary of the unit disc. Denote this region by $A$. This shows in particular that it has no intersection with ${\cal F}_{{t}}$.

As in the proof of Theorem \ref{Reachset1}, for any given  $t$,
$\lim_{\omega \rightarrow \infty} x(\omega)=1$ and $\lim_{\omega \rightarrow \infty} y(\omega)=0$.
Therefore ${\cal S}_t$ eventually belongs to this region.

Given a point $\left(x_{\omega}(t),y_{\omega}(t)\right)$ of the curve (\ref{ap-coordi1}) (\ref{ap-coordi2}), we denote by  $\phi(\omega)$ its phase if it belongs to ${\cal F}_t$, i.e. $\omega\in \left[ 0, \sqrt{\frac{\pi^2}{\bar{t}^2}- \gamma^2}\right]$, and we denote by  $\psi(\omega)$ its phase if it belongs to ${\cal S}_t$, i.e. $\omega\in \left(  \sqrt{\frac{\pi^2}{\bar{t}^2}- \gamma^2}, +\infty \right)$.

Using (\ref{ap-coordi1}) (\ref{ap-coordi2}), the phase of this initial point, i.e. when  $\omega=0$,
is $\phi(0)=0$ if $at=\gamma t \leq \frac{\pi}{2}$ and $\phi(0)=-\pi$ if
$\frac{\pi}{2} < at=\gamma t < \pi$. The portion of the curve (\ref{ap-coordi1}) (\ref{ap-coordi2}) belonging to ${\cal F}_t$, corresponds $at$ going from  $\gamma t$ to $\pi$.

To prove our statement, and show that, in fact, ${\cal S}_t$ never exits the region $A$,  we will  consider separately the two cases:
\begin{itemize}
\item  $\gamma t > \frac{\pi}{2}$, in which case the initial point corresponding to $\omega=0$ is on the negative side of the $x$ axis,  and $\phi(0)=-\pi$,
\item  $\gamma t \leq \frac{\pi}{2}$ in which case it is on the positive axis, and $\phi(0)=0$. The proof in this second case requires few extra elements.
\end{itemize}

\vs

{\bf { Case: $\gamma t > \frac{\pi}{2}$}}

\vs

We will prove that:
\begin{itemize}
\item[a.] The value of the phase, $\psi(\omega)$, of the corresponding point in ${\cal S}_t$ is always greater than the value of the phase (for ${\cal F}_t$) at the point corresponding to $\omega=\sqrt{\frac{\pi^2}{t^2}- \gamma^2}$ which is (cf.(\ref{fase})),
\be{phasemin}
\phi_0:=-\pi+\sqrt{\pi^2-\gamma^2 t^2}.
\ee
\item[b.]
This value $\phi_0$ is always greater  than all values of the phase, $\phi(\omega)$, of the corresponding   points in
${\cal F}_{{t}}$.\footnote{Here we take the `principal' value of the phase, i.e., $\psi \in [-\pi, \pi)$, and in this case, since we are in the negative $y$'s region $\psi \in [-\pi, 0]$.}
\end{itemize}

This gives that the phase for points on ${\cal S}_t$, $\psi(\omega)$,  is always greater than the phase for points on ${\cal F}_t$, $\phi(\omega)$, and therefore intersection cannot occur.

\vs

{\bf {a.}}
We first consider the function $\psi(\omega)$, where $\omega$ varies so that $at \in [\pi, \frac{3}{2}\pi]$ and then, for $k=2,3,...$, for
$at \in \left( ( 2k-1)\frac{\pi}{2}, ( 2k+1)\frac{\pi}{2} \right]$. If $at \in [\pi, \frac{3}{2}\pi)$ the phase (from the entry (1,1) in (\ref{soluzexpli3})) is
\be{faseagain}
\psi(\omega)=\omega t - \pi -\arctan\left(\frac{\omega}{a} \tan (at)\right).
\ee
Calculating $\frac{d \psi}{d \omega}$ we obtain, after some manipulations,
\be{dpdo}
\frac{d\psi}{d\omega}=t-\frac{a^2 \cos^2(at)}{\gamma^2 \cos^2(at)+\omega^2}\left(
\frac{\gamma^2}{a^3}\tan(at)+\frac{\omega^2t }{a^2 \cos^2(at)} \right).
\ee
To study the sign of $\frac{d\psi}{d\omega}$, we calculate $(\gamma^2 \cos^2(at) +\omega^2)a \frac{d \psi}{d \omega}$, whose sign is the same as the sign of $\frac{d \psi}{d \omega}$. We have
\be{dpdo2}
(\gamma^2 \cos^2(at) +\omega^2)a \frac{d \psi}{d \omega}=\gamma^2 \cos(at)(\cos(at)at-\sin(at)):=f(at).
\ee
For $at \in (\pi, \frac{3\pi}{2})$ this function is positive and then becomes negative with increasing $at$. It has one zero and $\lim_{at \rightarrow \frac{3 \pi}{2}} f(at)=0$. Therefore, to show that $\psi(\omega)$ is greater than the one corresponding to the value of $\omega=\sqrt{\frac{\pi^2}{t^2}-\gamma^2}$, i.e., the value where $at=\pi$, for all values of $at \in (\pi, \frac{3 \pi}{2}]$ is enough to show that the phase for $at=\frac{3 \pi}{2}$ is greater than the phase for $at=\pi$. This gives the inequality:
\be{Ltt3}
\phi_0=-\pi + \sqrt{\pi^2 -\gamma^2 t^2}<-\frac{3\pi}{2}+\sqrt{\frac{9 \pi^2}{4} -\gamma^2 t^2}.
\ee
For future use we prove the more general inequality
\be{moregen}
\phi_0=-\pi + \sqrt{\pi^2 -\gamma^2 t^2}< -\frac{(2k+1)\pi}{2}+ \sqrt{\frac{(2k+1)^2\pi^2}{4}-\gamma^2 t^2},
\ee
$k=1,2,...$ of which (\ref{Ltt3}) is a special case with $k=1$. Inequality (\ref{moregen}) is equivalent to
\be{moregenequiv}
(2k-1)\frac{\pi}{2} + \sqrt{\pi^2-\gamma^2 t^2} < \sqrt{\frac{(2k+1)^2\pi^2}{4}-\gamma^2 t^2}.
\ee
Squaring both terms and after some simplifications, we obtain,
\be{aftersqua}
(2k-1)^2\pi + 4 \pi + 4 (2k-1)\sqrt{\pi^2 - \gamma^2 t ^2} < (2 k +1)^2 \pi,
\ee
which, after some manipulations gives, the obviously true relation
\be{UIUp}
\sqrt{\pi^2 - \gamma^2 t^2} < \pi.
\ee

Generalizing (\ref{faseagain}), we define
\be{psik}
\omega t -k\pi -\arctan(\frac{\omega}{a} \tan(at)):=\psi_{k}(\omega),
\ee
when  $at \in \left( ( 2k-1)\frac{\pi}{2}, ( 2k+1)\frac{\pi}{2} \right)$, for $k=1,2,3,...$.
The general expression for the phase when $at \in [\frac{3 \pi}{2}, \infty)$ is given by the continuous function $\psi=\psi(\omega)$
\be{POLP}
\psi(\omega)=\psi_{k}(\omega),
\ee
when  $at \in \left( ( 2k-1)\frac{\pi}{2}, ( 2k+1)\frac{\pi}{2} \right)$, for $k=2,3,...$,
\be{Limiti}
\psi(\omega)=\lim_{\omega \rightarrow {\sqrt{(2k-1)^2 \frac{\pi^2}{4t^2}-\gamma^2}}^+} \psi_{k-1}(\omega)=\lim_{\omega \rightarrow \sqrt{(2k-1)^2 \frac{\pi^2}{4t^2}-\gamma^2}^-} \psi_{k}(\omega)=
\sqrt{(2k-1)^2\frac{\pi^2}{4}-\gamma^2t^2}-k \pi + \frac{\pi}{2},
\ee
in the points $at=\frac{(2k-1)\pi}{2}$, $k=2,3,...$.\footnote{We remark that this is the `principal' value of the phase, that is, the one with value in the interval $(-\pi, 0)$. To see this, notice that at the first endpoint of the interval $[(2k-1)\frac{\pi}{2}, (2 k+1)\frac{\pi}{2}]$, we have $\psi(\omega)$ given by $\sqrt{(2k-1)^2\frac{\pi^2}{4}-\gamma^2t^2}-k \pi + \frac{\pi}{2}$ as in (\ref{Limiti}) and we can see $
0 >\sqrt{(2k-1)^2\frac{\pi^2}{4}-\gamma^2t^2}-k \pi + \frac{\pi}{2} > -\pi.$
The right inequality is equivalent to $
\sqrt{(2k-1)^2\frac{\pi^2}{4}-\gamma^2 t^2}> k\pi  -\frac{3 \pi}{2},
$
which squaring both terms and after simplifications leads to the true inequality
$
-\gamma^2 t^2 >- 2(k-1)\pi^2.
$
The left inequality follows similarly. The fact that the phase $\psi=\psi(\omega)$ remains in this range is a consequence of the considerations that follow.}
In the interior of the interval where $at \in \left[ ( 2k-1)\frac{\pi}{2}, ( 2k+1)\frac{\pi}{2} \right]$, for $k=2,3,...$,  the derivative of the function $\psi(\omega)$ is still given by (\ref{dpdo}) and it is positive, then, it becomes negative and then it tends again to zero as $at \rightarrow \frac{(2k+1)\pi}{2}$. This means that $\psi(\omega)$ is increasing, it reaches a maximum and then it decreases to a minimum. The maximum cannot be greater than zero because by continuity the curve (\ref{ap-coordi1}) (\ref{ap-coordi2}) will have to cross the $x$ axis and we have seen that this happens only when $\omega=0$. This ensures that $\psi(\omega)<0$. The minimum the function tends to is the value of the function at  $at=(2k+1)\frac{\pi}{2}$, i.e., $\sqrt{(2k+1)^2 \frac{\pi^2}{4}-\gamma^2 t^2}-(2k+1)\frac{\pi}{2}$, which is greater than $\phi_0:=-\pi + \sqrt{\pi^2 -\gamma^2 t^2}$ (and therefore of $-\pi$) as it was proved in (\ref{moregen}).

\vs

{\bf {b.}}  The phase $\phi(\omega)$ of a point  belonging to ${\cal F}_t$, which corresponds to $at$ going from  $\gamma t$ to $\pi$, can be written, since we have $\gamma t>\frac{\pi}{2}$:
\be{aphi}
\phi(\omega)=\omega t - \pi -\arctan\left(\frac{\omega}{a} \tan (at)\right).
\ee
This expression is the same as the one given in equation (\ref{faseagain}). Thus if we take the derivative with respect to $\omega$, we end up with the same $f(at)$ function of equation (\ref{dpdo2}). This function is positive  for $(\frac{\pi}{2}<)\gamma t<at<\pi$, thus the phase is always increasing so $\phi(\omega)\leq \phi_0$.

\vs

{\bf { Case: $\gamma t \leq \frac{\pi}{2}$}}

\vs

In this case, the previous argument has to be modified with some extra elements since the phase $\phi(\omega)$ of the points of ${\cal F}_t$, is not always increasing. Here we have:
\be{P1}
\phi(\omega)=\omega t -\arctan\left(\frac{\omega}{a} \tan(at)\right),
\ee
if $at \in [\gamma t, \frac{\pi}{2})$. Then $\phi(\omega)$ is given by equation (\ref{aphi}), for
$at\in [\frac{\pi}{2},\pi]$. The function $\phi(\omega)$ has a minimum for $at =\frac{\pi}{2}$ and then increases for $at\in (\frac{\pi}{2}, \pi]$. Moreover, we have that $\phi(\frac{\pi}{2})=\phi_0$ in (\ref{phasemin}).

The phase $\psi(\omega)$ of the points in ${\cal{S}}_t$, is again given by equations
  (\ref{psik}), (\ref{POLP}), (\ref{Limiti}), for $at\in (\pi, \infty)$.
 Proceeding as in the previous case (part a.), we show that the phase for the points of ${\cal S}_t$ has to be always greater than the value $\phi_0$, and therefore greater of all points on ${\cal F}_t$ for
 $at\in (\frac{\pi}{2}, \pi]$. Therefore if  an intersection occurs it has to occur in the first part of the curve ${\cal F}_t$, the one corresponding to values of $at \in [\gamma t, \frac{\pi}{2})$. Now consider the square radius $r^2$ as a function of $at$, i.e.,
\be{sqrrad}
r^2=1-\frac{\gamma^2 t^2}{a^2 t^2} \sin^2(at).
\ee
For $at \in [\gamma t, \pi]$ this function is increasing as a function of $at$. Therefore the radius on the first part of ${\cal F}_t$ is $\leq$ than the value of this function at $at=\frac{\pi}{2}$, which is,
\be{yyio}
r_0^2=1-\frac{4 \gamma^2 t^2}{\pi^2}.
\ee
Now for $at \in [\pi,  \frac{3\pi}{2}]$ the square radius function is decreasing, it has a minimum and then it is increasing again. A computation of the derivative with respect to $at$ shows that the minimum is at the point $at=z_1 \in (\pi, \frac{3 \pi}{2})$ such that $\tan(z_1)=z_1$. At this point, writing $\sin^2(z_1)=\frac{\tan^2(z_1)}{1+\tan^2(z_1)}$, we can write the radius in (\ref{sqrrad}) as
\be{r2}
r_1^2=1-\frac{\gamma^2 t ^2}{1+z_1^2}.
\ee
Comparing with (\ref{yyio}) using the fact that $z_1 > \pi$, we obtain $r_1 > r_0$, i.e., the minimum radius for $at \in [\pi, \frac{3 \pi}{2}]$ is greater than the value $r_0$.

This argument extends to vales of $at \in [(2k-1) \frac{\pi}{2}, (2k+1) \frac{\pi}{2}]$, for any $k \geq 2$. In these intervals, the radius square function is increasing, then decreasing, then increasing again. The minimum is obtained for a value $z_k \in ((2k-1) \frac{\pi}{2}, (2k+1) \frac{\pi}{2})$ such that $\tan(z_k)=z_k$, and for the corresponding radius $r_k$, we have:
\[
r_k>r_1>r_0,
\]
so also in this case the curves ${\cal F}_{t}$ and  ${\cal S}_t$ do not intersect.

\end{document}